\documentclass[10pt,conference]{IEEEtran}
\usepackage{amsfonts}
\usepackage{amssymb}

\usepackage{algorithmic}
\usepackage{graphicx}
\usepackage{textcomp}
\usepackage{hyperref}
\usepackage{multirow}
\usepackage{caption}
\usepackage{twemojis}
\usepackage{booktabs}
\usepackage[T1]{fontenc}
\usepackage{tabularx}
\usepackage{amsmath}

\usepackage[table]{xcolor}
\def\BibTeX{{\rm B\kern-.05em{\sc i\kern-.025em b}\kern-.08em
    T\kern-.1667em\lower.7ex\hbox{E}\kern-.125emX}}

\newcommand\todo[1]{\textcolor{red}{#1}}
\newcommand\changed[1]{\textcolor{black}{#1}}
\newcommand\changedd[1]{\textcolor{black}{#1}}

\begin{document}

\title{\changed{Psychological Safety Framework in Pull-based Open Source Projects}}

\author{\IEEEauthorblockN{Emeralda Sesari}
\IEEEauthorblockA{
\textit{University of Groningen}\\
Netherlands \\
e.g.sesari@rug.nl}
\and
\IEEEauthorblockN{Federica Sarro}
\IEEEauthorblockA{
\textit{University College London}\\
United Kingdom \\
f.sarro@ucl.ac.uk}
\and
\IEEEauthorblockN{Ayushi Rastogi}
\IEEEauthorblockA{
\textit{University of Groningen}\\
Netherlands \\
a.rastogi@rug.nl}
}

\maketitle

\begin{abstract}
Psychological safety (PS) refers to the belief that team members can speak up, ask questions, and make mistakes without fear of negative consequences. Although PS has been studied in traditional software teams, less is known about how it may appear in pull-based open-source software development, where contributors are self-directed and often collaborate voluntarily.
This paper introduces a theory-informed framework for understanding how PS may be reflected in pull request (PR) interactions. We investigate the framework through a mixed-methods approach combining repository-based analysis and a survey of OSS contributors. First, we constructed a PS index using nine observable PR interaction indicators extracted from 60,685 PRs across 26 GitHub repositories. Second, we surveyed 57 OSS contributors about their perceived PS and experiences of corresponding PR interactions.
All nine surveyed interaction indicators were significantly and positively associated with perceived PS. However, the repository-based index was not significantly associated with survey-reported PS. Contributors' open-ended responses further highlighted the importance of feedback quality, opportunities to participate, and how PR outcomes are interpreted.
Together, our framework and mixed-methods findings provide initial insights into how PS may be reflected in PR interactions, highlighting the importance of both PR interaction patterns and contributors' experiences.
\end{abstract}


\section{Introduction}
Software Engineering (SE) is a highly team-intensive discipline that requires effective collaboration, communication, and coordination among developers to successfully build and maintain software systems. Studies have shown that factors such as team diversity, cohesion, and human factors impact software development outcomes \cite{pieterse2006software,Capretz2014}. One of the key elements that fosters such teamwork is psychological safety (PS). Psychological safety is a shared belief held by members of a team that the team is safe for interpersonal risk-taking \cite{edmondson1999psychological}. It allows individuals to speak up, ask questions, admit mistakes, and offer new ideas without fear of negative consequences. Research has shown that PS is a key factor in high-performing teams \cite{edmondson1999psychological}.

Google’s Project Aristotle examined 115 engineering teams and 65 sales teams to determine what makes a team effective \cite{nytimesWhatGoogle}. The results revealed that teams with strong PS are more effective, as members feel comfortable sharing ideas, taking risks, and collaborating openly. It was also found that even highly skilled employees need a psychologically safe environment to contribute their best work. PS allows individuals to bring their entire self to work, to allow honest discussions about work challenges, frustrations, and uncertainties. GitLab emphasizes the negative effects of punishment culture, blame, and poor communication, which can reduce trust and collaboratio \cite{gitlabPsychologicalSafety}. PS has also been linked to high team performance in software teams \cite{nytimesWhatGoogle, Lenberg2018}.

Despite these insights, much of the empirical research on PS in SE has focused on software teams in organizational settings \cite{Santana2023, Santana2024, Lenberg2018, Alami2024, Kakar_2018}. 
However, software development also takes place in open-source software (OSS) communities, which might greatly differ from industrial workplaces. 
OSS contributors span diverse time zones, cultures, and experience levels, and they participate voluntarily or as part of their professional work \cite{vasilescu2015perceptions}. Because participation is self-directed and non-contractual, contributors are especially sensitive to the social dynamics they encounter. Prior studies have shown that motivation \cite{fang2009understanding,roberts2006understanding,hertel2003motivation}, task characteristics \cite{lin2017developer,schilling2012will}, social capital \cite{Qiu2019}, and experiences of rejection \cite{jiang2013will,hellendoorn2015will,padhye2014study,tao2014writing,gousios2014exploratory} shape contributor engagement. Given this, psychological safety, the sense that it is safe to ask questions and make mistakes, may be particularly important in OSS, where contributors often decide to stay or leave based on how they are treated and whether their work is valued.

In OSS development, pull requests (PRs) are central to collaboration. PRs are not only technical data but also a source of social interaction, where contributors receive feedback, negotiate changes, and seek acceptance from project maintainers \cite{alami2021pull}. These interactions can be constructive, supportive, or discouraged, and offer insight into how teams communicate \cite{li2017they}. Although such traces do not directly capture contributors’ internal feelings of safety, they can help identify interaction patterns that may be related to PS in pull-based OSS development.

\changedd{In this study, we propose a framework grounded in PS theory to understand how PS may be reflected in PR interactions. \footnote{A theoretical framework provides the structure for understanding a phenomenon, particularly when applying a theory from another field, and helps clarify the context and relevance of the study~\cite{lynham2002general,runeson2012case}} We investigate the framework through a mixed-methods approach combining repository-based analysis and a survey of OSS contributors. The repository-based analysis constructs a PS index using selected observable PR interaction patterns, while the survey examines contributors' experiences of corresponding PR interactions and how these relate to their perceived PS. 
By considering both approaches together, we examine what each reveals about how PS may be reflected in PR interactions and where their findings differ.}

\changedd{The contributions of this paper are threefold: (1) we present a theory-informed framework for understanding how PS may be reflected in PR interactions within OSS projects; (2) we construct an index using selected observable PR interaction indicators; and (3) we provide empirical insights into contributors' reported experiences of PR interactions and their relationship with perceived PS, alongside comparisons with repository-based findings.}




\section{Background and Related Work}\label{sec:II}
\subsection{Psychological Safety}
Originating from organizational behavior field, the concept of PS was introduced by Amy Edmondson in 1999 through her study of clinical teams \cite{edmondson1999psychological}. She highlighted the importance of a shared belief that a team is safe for interpersonal risk-taking \cite{edmondson1999psychological}. Edmondson identified key indicators of PS, including whether team members feel comfortable raising concerns, taking risks, and seeking help without fear of embarrassment or punishment \cite{edmondson1999psychological}. The following are the survey items developed by Edmondson to measure team PS.

\begin{enumerate}
    \item If you make a mistake on this team, it is often held against you.
    \item Members of this team are able to bring up problems and tough issues.
    \item People on this team sometimes reject others for being different.
    \item It is safe to take a risk on this team.
    \item It is difficult to ask other members of this team for help.
    \item No one on this team would deliberately act in a way that undermines my efforts.
    \item Working with members of this team, my unique skills and talents are valued and utilized.
\end{enumerate}

Subsequent research, such as Google's Project Aristotle, found PS as a key determinant of team effectiveness \cite{nytimesWhatGoogle}. After studying 180 teams comprising 115 engineering teams and 65 sales teams, Google found that team composition, such as shared hobbies, personalities, or management styles, did not predict team effectiveness. Instead, Google's study identified five key factors of high-performing teams: impact, meaning, structure \& clarity, dependability, and PS, with PS being the strongest predictor of team success. 

\subsection{Psychological Safety in Software Teams}
In recent years, researchers have examined PS impact in various SE contexts, from Stack Exchange discussions to Agile team dynamics. Previous studies \cite{Santana2023,Santana2024,Lenberg2018,Alami2024, Kakar_2018} together observed that PS is a vital component of successful software development teams, which foster an environment where developers feel comfortable sharing ideas, asking questions, and embracing innovation.

Santana et al. investigated PS in SE by analyzing discussions on Stack Exchange \cite{Santana2023}. Their findings suggest that a psychologically safe environment enables developers to ask questions and share knowledge more freely, leading to enhanced learning and problem-solving efficiency~\cite{Santana2023}. Additionally, they identified seven potential indicators of PS in software development teams through various interpersonal scenarios (see first row in Table~\ref{tab:PSinST}). 
Further studies by Santana et al. highlight the significance of PS in software workplaces \cite{Santana2024}. Their research emphasizes that teams with higher levels of PS exhibit improved communication, reduced stress, and greater openness to innovation, which are important for maintaining productivity and software quality~\cite{Santana2024}. They also shared some takeaways for software teams to foster a psychologically safe workspace, such as for software project leaders to ensure that the team is comfortable admitting errors. They observed that in a psychologically safe workplace, there are seven points that should not be challenging (see second row in Table~\ref{tab:PSinST}).

The role of PS in Agile teams has also been a subject of interest. \changed{Tkalich et al.~\cite{tkalich2022} examined what happens to PS when software teams move to remote work. They found that PS remains important in settings where collaboration is mediated by digital communication.} Alami et al. conducted interviews with 20 practitioners and surveyed 423 participants \cite{Alami2024}. They observed that psychologically safe Agile software development teams seem to capitalize on six main quality-related behaviors (see last row in Table~\ref{tab:PSinST}).
Additionally, Lenberg and Feldt explored the relationship between PS and norm clarity within SE teams \cite{Lenberg2018}. They collected industry survey data from practitioners (N = 217) in 38 development teams working for five different organizations. The result of multiple linear regression analyses indicates that both PS and team norm clarity predict team members’ self-assessed performance and job satisfaction.

Moreover, Kakar examined the interplay between PS, team cohesion, and knowledge sharing in software development projects \cite{Kakar_2018}. 
They found that high levels of team cohesion can leverage the positive impact of PS on knowledge sharing, ultimately enhancing collaboration and team efficiency~\cite{Kakar_2018}.

\changed{More recently, Sanchez-Gordon et al.~\cite{sanchezgordon2026voice} examined PS together with speaking up and staying silent behaviors among software professionals. They found that PS was more strongly associated with reduced silence than with increased speaking up, while silence was associated with withdrawal and speaking up was associated with work performance. }

Most research on PS in software teams has focused on company environments using qualitative methods such as interviews and surveys. However, software development also occurs in OSS communities. To explore how PS operates in OSS settings, the next section focuses on developing a framework grounded in PS theory to explain what PS can look like in this unique context. 

\begin{table}[]
\footnotesize
\captionsetup{font=footnotesize}
\caption{Indicators of Psychological Safety in Software Teams}
\resizebox{\columnwidth}{!}{%
\begin{tabular}{c|l}
\hline
Source                          & \multicolumn{1}{c}{PS Indicators}                          \\ \hline
\multirow{7}{*}{Santana et al. \cite{Santana2023}}                  & Communication of opinions on work issues                                     \\
                                & Valuation of expressed opinion                                               \\
                                & Recommendations/ideas for new projects or changes in procedures              \\
                                & Speaking up about other people's mistakes                                    \\
                                & Security in answering questions or solving uncertainties in relation to work \\
                                & Existence of actual attempts to share information                            \\
                                & Encouragement and support to take on new tasks or learn something new        \\ \hline
\multirow{7}{*}{Santana et al. \cite{Santana2024}} & Disagreeing with suggestions or ideas is not challenging                     \\
                                & Identifying problems is not challenging                                      \\
                                & Recommending changes is not challenging                                      \\
                                & Sharing negative feedback is not challenging                                 \\
                                & Seeking help is not challenging                                              \\
                                & Drawing attention to errors is not challenging                               \\
                                & Admitting errors is not challenging                                          \\ \hline
\multirow{6}{*}{Alami et al. \cite{Alami2024}}   & Speaking up about software quality problems                                  \\
                                & Admitting software quality mistakes                                          \\
                                & Learning from mistakes                                                       \\
                                & Helping each other to enhance quality                                        \\
                                & Collective problem-solving to solve quality issues                           \\
                                & Taking initiatives to promote better software quality                        \\ \hline
\end{tabular}
}
\label{tab:PSinST}
\end{table}

\section{Psychological Safety in Pull Request Interactions}\label{sec:III}
OSS development brings together contributors from different time zones, backgrounds, and levels of experience, many of whom participate either voluntarily or as part of their professional responsibilities \cite{vasilescu2015perceptions}. In OSS, contributors are especially attuned to the social dynamics they encounter such as motivation\cite{fang2009understanding,roberts2006understanding,hertel2003motivation}, task characteristics\cite{lin2017developer,schilling2012will}, social capital \cite{Qiu2019}, and experiences of rejection\cite{jiang2013will,hellendoorn2015will,padhye2014study,tao2014writing,gousios2014exploratory}.

A central aspect of OSS collaboration is the pull-based development model, in which contributors submit changes via PR. Contributors share their work for feedback, have conversations with others, and deal with different styles of communication and decision-making. 
 Trinkenreich et al. \cite{Trinkenreich2025} found that many contributors, especially those from underrepresented groups, experience significant interpersonal challenges in OSS communities, including hostile communication, microaggressions, lack of support, and unequal recognition. These issues often surface during PR discussions, which shape contributors’ sense of belonging and willingness to engage~\cite{Trinkenreich2025}. 
 

Given the central role of PR in shaping contributor experience, we build on this foundation by proposing that PS conditions may be reflected in the interaction patterns that unfold in PRs. Inspired by German et al.'s work on applying fairness theory in modern code reviews~\cite{german2018my}, we draw on PS theory to construct a framework for understanding how PR interactions may reflect conditions that support or undermine PS. 
Applying PS theory to PR interactions has the potential to enhance our understanding of how team dynamics influence contributor behavior, such as sense of belonging~\cite{Trinkenreich2025}, willingness to participate~\cite{Trinkenreich2025}, and ultimately support sustained participation in OSS communities~\cite{Qiu2019}.

\subsection{Proposed Framework}

\changedd{We developed the framework by drawing on established PS theory \cite{edmondson1999psychological} and prior studies of PS-related behaviors in software teams \cite{Santana2023,Santana2024,Alami2024}, and considering how these behaviors may manifest in PR interactions. The resulting framework identifies seven theory-informed aspects relevant to PR interactions in OSS, providing a foundation of how PS may look like in PR interactions.} In this section, we present a framework we developed to illustrate how PS can be reflected in PR interactions.

\textbf{Giving and Receiving Feedback.}
In a psychologically safe environment, open discussions take place around PRs. Contributors feel comfortable giving and receiving feedback, which often leads to richer conversations. A higher volume of discussion can suggest that contributors feel safe expressing their thoughts, asking questions, and responding to others \cite{ferguson2024no}.

This behavior reflects several key aspects of PS. Edmondson defines PS as a belief that one can speak up about work-related issues without fear of negative consequences \cite{edmondson1999psychological}. Similarly, Santana et al. \cite{Santana2023} observed that in psychologically safe teams, software team members openly communicate their opinions, highlight others’ mistakes, ask questions, and share information freely. Santana et al. \cite{Santana2024} also emphasized that in such environments, disagreeing with suggestions, recommending changes, sharing negative feedback, and admitting errors are seen as routine and constructive aspects of collaboration. Finally, Alami et al. \cite{Alami2024} noted that in such environments, team members help each other to enhance software quality and engage in collective problem-solving.

\textbf{Taking Risks Without Fear.}
In a psychologically safe team, contributors feel confident taking risks. When a PR is rejected, a psychologically safe environment allows contributors to reopen the discussion, revise their work, and try again without feeling embarrassed or discouraged. Mistakes are seen as a natural part of the development process, and correcting them is encouraged rather than penalized.

This behavior aligns with Edmondson’s belief that “it is safe to take a risk on this team” and that “no one on this team would deliberately act in a way that undermines one’s efforts” \cite{edmondson1999psychological}. It also echoes the findings by Santana et al. \cite{Santana2023}, who emphasize that psychologically safe teams provide encouragement and support to take on new tasks and learn new skills. Their later study \cite{Santana2024} adds that admitting errors is seen as a normal and accepted behavior. Alami et al. \cite{Alami2024} similarly observed that in such teams, developers are not afraid to acknowledge mistakes and treat them as opportunities for learning and growth.

\textbf{Active Participation from Contributors and Reviewers.}
In a psychologically safe project, both contributors and reviewers engage in PR discussions, rather than one side dominating the conversation. This exchange allows contributors to explain their decisions, seek clarification, and respond to suggestions, while reviewers provide respectful and constructive feedback \cite{golzadeh2019effect}. Most PRs contain some form of discussion, which shows how central communication is to the review process \cite{golzadeh2019effect}. When both parties participate, it suggests a shared commitment to collaboration and learning.

Such interactions signal PS: contributors feel their voices matter, and reviewers show respect for others' efforts by taking time to explain their reasoning and invite further dialogue. This aligns with Edmondson’s notion that psychologically safe teams value each member’s contributions and do not undermine individual efforts \cite{edmondson1999psychological}. Similarly, prior work highlights that contributors feel secure asking questions, expressing opinions, and engaging in mutual efforts to improve software when PS is present \cite{Santana2023,Alami2024}.

\textbf{Conflict as an Opportunity for Growth.} In pull-based development, merge conflicts are common \cite{gousios2014exploratory}. They occur when changes in a PR interfere with newer updates in the project’s main branch. GitHub flags these conflicts, and while either the contributor or a core team member can resolve them, the prevailing etiquette expects the contributor to bring the PR back into a non-conflicting state \cite{gousios2014exploratory}. This process requires not just technical effort but also communication, collaboration, and mutual understanding between contributors and reviewers.

In a psychologically safe environment, conflict is not viewed as a threat but as a normal and potentially productive part of collaboration. Teams with high levels of PS are more likely to see conflict as an opportunity for growth rather than a barrier to progress \cite{o2018optimizing}. This reflects Edmondson’s PS dimension that “people on this team do not reject others for being different” \cite{edmondson1999psychological}, and aligns with Santana et al.’s findings that disagreeing with suggestions or ideas is not seen as challenging in psychologically safe teams \cite{Santana2024}.

\textbf{Team Members Actively Raising Issues.}
In a psychologically safe team, members feel comfortable raising concerns and engaging in discussions. In PR interactions, this is reflected when contributors are actively involved in explaining or defending their code \cite{golzadeh2019effect, kononenko2018studying}, integrators respond with collaborative and respectful feedback \cite{golzadeh2019effect}, and other project members contribute to the conversation \cite{golzadeh2019effect}. When review discussions include more than just the contributor and a single reviewer, it also signals an inclusive and communicative environment \cite{kononenko2018studying}.

These patterns indicate a setting where difficult issues can be surfaced and worked through constructively, which is a core aspect of PS \cite{edmondson1999psychological}. Alami et al. \cite{Alami2024} similarly found that psychologically safe teams support each other and engage in collective problem-solving, behaviors that are often visible in how PR conversations unfold. This is also consistent with recent SE research on voice and silence, which found that PS is associated with reduced silence among software professionals~\cite{sanchezgordon2026voice}.

\textbf{Asking for Input.}
In a psychologically safe environment, contributors feel empowered to involve others directly in the review process. One way this manifests in PR interactions is through the use of direct mentions to tag specific individuals. Prior work also suggests that directly tagging someone in a pull request can significantly reduce the time to review, likely because it creates a sense of social responsibility or obligation for the tagged user \cite{yu2016determinants}. This proactive communication behavior reflects a level of interpersonal trust and confidence that is more common in psychologically safe teams. 

Ferguson et al. \cite{ferguson2024no} found that teams with higher PS engaged more frequently in this behavior, using mentions as part of their learning-oriented interactions. Tagging others signals that contributors are not hesitant to ask for input, clarifications, or support, which indicates a culture where requesting help is accepted and even encouraged.

\textbf{Familiarity Behavior.} Familiarity behavior such as using emotional expressions is a hallmark of socially close and trusting teams. In conversations, one subtle but revealing expression of familiarity is the use of emojis \cite{ferguson2024no}. Emojis signal emotional tone, solidarity, and casual rapport, which are more likely to occur in psychologically safe environments where people feel comfortable being themselves \cite{ferguson2024no}. Ferguson et al. found that teams with higher PS used significantly more emoji reactions than teams with lower PS \cite{ferguson2024no}. 

In the context of PRs, developers pay close attention to emojis in PR comments, interpreting them as part of the overall sentiment even more than textual tone in some cases~\cite{park2021assessing}. Emojis also help reduce comment clutter by allowing reviewers to express approval or support through simple reactions like \twemoji{thumbsup} or \twemoji{1f389}, especially in teams with established relationships~\cite{wang2023more}.

This proposed framework offers a theory-informed lens for identifying PS conditions that may be visible in PR interactions. It should not be understood as the only way to conceptualise PS in OSS. 
The empirical study that follows examines the proposed framework through both observable PR interaction patterns and contributors' reported experiences.

\section{\changedd{Research Methodology}}
\subsection{Research Design}
To investigate how PS framework may be reflected in PR interactions and relevant to contributors' experience, we adopted a mixed-methods approach combining repository-based analysis with a survey of OSS contributors. Both approaches were guided by the proposed framework, but examine different perspectives of PS.
First, the repository-based analysis examines patterns of PR interactions using GitHub data. We examined the framework through observable PR interactions that may reflect PS. Second, the survey examines contributors' perceived PS and their experiences of the corresponding PR interactions. The survey also includes open-ended responses to capture contributors' experiences that may not be reflected in the framework. 
Following the approach of integrating mixed-methods findings through interpretation and reporting \cite{fetters2013integration}, we considered the findings together by examining the association between repository-based indices and survey-reported PS, and interpreting the quantitative results alongside contributors' open-ended responses. This allowed us to examine both the relationships and differences between observable PR interaction patterns and contributors' experiences.

\subsection{Dataset and Repository Selection}\label{reposelection}
For the repository-based analysis, we used the dataset provided by Zhang et al. \cite{zhang2020shoulders}, which is the largest dataset for pull-based development research to date. This dataset contains 11,230 OSS projects, 
including 96 metrics and 3,347,937 pull requests. 
It is the most comprehensive and largest dataset toward a complete picture of pull-based development research \cite{zhang2020shoulders}. Their feature selection is based on Gousios et al.'s dataset \cite{gousios2014dataset}, as well as studies on pull request development.
These features broadly fall into three categories: those related to the contributor, the project, and the pull request, though some features lie at their intersection. 

This makes the dataset suitable for our study as it contains the information needed to examine the PS framework.


To ensure a relevant dataset for our analysis, we applied a filtering process to select repositories with active engagement. Prior work suggests that starring can be a useful proxy for project engagement, as popular repositories tend to attract more contributors and sustained development activity \cite{begel2013social,borges2016understanding,borges2018s}.

However, recent research has highlighted concerns regarding fake stars on GitHub, where artificially inflated popularity has been linked to scams, malware, and deceptive ranking tactics \cite{he202445millionsuspectedfake}. To mitigate this risk, we initially selected the 200 most starred repositories but further refined the repositories by removing repositories that did not primarily focus on active software development contributions. Specifically, we excluded:
\begin{itemize}
    \item code learning (e.g., coding challenges, tutorials);
    \item resource lists (e.g., curated lists of tools, frameworks, or libraries);
    \item educational content (e.g., repositories intended for teaching rather than active software development);
    \item non-English repositories, as language barriers limit our understanding about the context of the repository;
    \item documentation-only repositories that do not involve active software development.
\end{itemize}

After applying these criteria, we retained 26 repositories, comprising a total of 60,685 PRs for repository-based analysis (see Sec. \ref{sec:operationalize}). We also recruited survey participants from these same repositories (see further in Sec.~\ref{sec:survey}). This allowed us to examine observable PR interactions and contributors' reported experiences within a common set of projects. The complete list of selected repositories, along with the PR distribution, is presented in our supplementary material~\cite{figshare}.

\section{Repository-Based Analysis}\label{sec:operationalize}
To examine the proposed framework using GitHub data, we identified observable PR interaction indicators and combined them into an index. This section describes how the indicators were selected and how the index was constructed.

\subsection{Mapping the Framework to Observable PR Indicators}


To identify potential variables for PS in OSS, we mapped the features available in the Zhang et al. dataset \cite{zhang2020shoulders} to our PS framework, selecting those that could act as representative variables. For instance, to capture the framework item \textit{active participation from contributors and reviewers}, we use the variable \texttt{has\_exchange}, which indicates whether a PR discussion contains a back-and-forth exchange between participants. When \texttt{has\_exchange = 1}, the corresponding indicator \texttt{PS\_has\_exchange} is assigned a value of 1 and contributes one point to the PR-level PS index. 
Another example is the sixth framework item, asking for input, which suggests that contributors feel comfortable involving others directly in the review process. This is reflected in the presence of direct mentions within PR comments, captured by the \texttt{at\_tag} variable.

\begin{table*}[]
\centering
\captionsetup{font=footnotesize}
\caption{List of PS Observable Variables Aligned with PS Framework}
\footnotesize
\label{tab:proxies}
\begin{tabular}{llll}
\hline
\multicolumn{1}{c}{PS Framework Item}  & \multicolumn{1}{c}{Variable} & \multicolumn{1}{c}{Description}                                                                                                                                                                                                                     \\ \hline
Giving and receiving feedback  & pr\_comment\_num            & \# of comments in a PR                                                                                                                                                       \\
\rowcolor[HTML]{EFEFEF}Taking risks without fear  & \todo{reopen\_num}               & How many times a PR reopened                                                                                                                             \\
Active participation from contributors and reviewers  &has\_exchange               & Has contributor and integrator comments? yes/no                                                                                                                                                                     \\
\rowcolor[HTML]{EFEFEF}Conflict as an opportunity for growth   & \todo{comment\_conflict}           & Keyword conflict exists in the comment          \\
\multirow{6}{*}{Team members actively raising issues}  & contrib\_comment            & Has a contributor comment? yes/no                                                                                                                                                            \\
&num\_comments\_con          & \# of contributor comment                                                                                                                                                                                                                                                        \\
&inte\_comment               & Has an integrator comment? yes/no                                                                                                                                                                                                                                                 \\
&reviewer\_comment           & Has a reviewer comment? yes/no                                                                                                                                                                                                                                                    \\
&other\_comment              & Has a non contributor/core team comment? yes/no                                                                                                                                                                                                                                   \\
&num\_participant            & \# of participants in pr comment                                                                                                                                                                                                                                                  \\
\rowcolor[HTML]{EFEFEF}Asking for input & at\_tag                     & Mentions exist? yes/no                                                                                                                                               \\
Familiarity behavior  & \todo{emoji\_count}                & \# of emojis in the comment                                                                                                                                                                                                                                                                                          \\\hline
\end{tabular}
\end{table*}


In addition to the variables identified from Zhang et al. dataset \cite{zhang2020shoulders}, we introduced an additional variable, \texttt{emoji\_count} to represent the seventh item of our PS framework \textit{familiarity behavior}. To compute this variable, we developed an R script that retrieves PR comments and counts the number of emojis present in the text, using a reference emoji dataset\footnote{\url{https://www.kaggle.com/datasets/subinium/emojiimage-dataset/data}}. The descriptive statistics for all variables are presented in our supplementary material \cite{figshare}. 
A detailed table listing our PS framework, the variables representing, and their descriptions is provided in Table \ref{tab:proxies}. 

To reduce the impact of extreme values in the candidate PS observable variables, we examined the distribution of continuous variables by computing their skewness. Using the \texttt{skewness()} function from the \texttt{e1071} package in R, we found that several variables, namely \texttt{pr\_comment\_num}, \texttt{reopen\_num}, \texttt{num\_comments\_con}, \texttt{num\_participants}, and \texttt{emoji\_count}, exhibited high positive skewness. To address this, we applied a log-transformation to these variables and re-examined their distributions. The transformation reduced skewness for \texttt{pr\_comment\_num}, \texttt{num\_comments\_con}, and \texttt{num\_participants}. However, \texttt{reopen\_num} and \texttt{emoji\_count} remained highly skewed after transformation. Given this, we excluded these two variables from the final empirical index to avoid unstable measurement and preserve interpretability~\cite{heinze2002}.

We also reviewed the distribution of binary variables. While most binary predictors showed relatively balanced distributions, the variable \texttt{comment\_conflict} was found to be extremely imbalanced, with only 1.8\% of instances taking the value 1. Due to the risk of unreliable statistical estimation~\cite{king2001logistic, abd2016imbalance}, this variable was excluded from the final set of PS observable variables. In total, we retained 9 variables (highlighted in black font in Table \ref{tab:proxies}) for further analysis.

\subsection{Constructing Psychological Safety Index}
\label{sec:constructing}
\begin{table*}[t]
\centering
\captionsetup{font=footnotesize}
\caption{Scoring rules for the PR-level psychological safety index. Each indicator is assigned a value of 1 when the corresponding condition is satisfied and 0 otherwise.}
\label{tab:condition}
\footnotesize
\begin{tabular}{p{4.6cm} p{7.5cm} p{4.5cm}}
\toprule
\multicolumn{1}{c}{PR-level condition} & 
\multicolumn{1}{c}{What it reflects} & 
\multicolumn{1}{c}{Indicator value} \\
\midrule
\texttt{pr\_comment\_num} is high 
& High volume of PR discussion 
& \texttt{PS\_pr\_comment\_num = 1} \\

\texttt{has\_exchange = 1} 
& Bidirectional communication in the PR thread 
& \texttt{PS\_has\_exchange = 1} \\

\texttt{contrib\_comment = 1} 
& Contributor participates in the discussion 
& \texttt{PS\_contrib\_comment = 1} \\

\texttt{num\_comments\_con} is high 
& Deeper contributor involvement in the discussion 
& \texttt{PS\_num\_comments\_con = 1} \\

\texttt{inte\_comment = 1} 
& Integrator is visibly involved in the PR discussion 
& \texttt{PS\_inte\_comment = 1} \\

\texttt{reviewer\_comment = 1} 
& Reviewer is visibly involved in the PR discussion 
& \texttt{PS\_reviewer\_comment = 1} \\

\texttt{other\_comment = 1} 
& Other community members join the discussion 
& \texttt{PS\_other\_comment = 1} \\

\texttt{num\_participants} is high 
& Broader participation in the PR discussion 
& \texttt{PS\_num\_participant = 1} \\

\texttt{at\_tag = 1} 
& Direct social acknowledgement or invitation to participate 
& \texttt{PS\_at\_tag = 1} \\
\bottomrule
\end{tabular}
\label{tab:condition}
\end{table*}
\changed{To examine the PS framework, we constructed the index at three levels: PR, contributor, and repository. Let $PS_j$ denote the PR-level PS score of PR $j$, $PS_{c,r}$ the contributor-level PS score of contributor $c$ in repository $r$, and $PS_r$ the repository-level PS score.}
\changed{At the PR level, each indicator in Table~\ref{tab:condition} is coded as a binary value. The PR-level PS score is the sum of these indicators:}

\changed{\begin{equation}
PS_j = \sum_{k=1}^{9} I_{j,k},
\end{equation}}

\changed{where $I_{j,k}$ is the binary value of indicator $k$ for PR $j$. Thus, higher values indicate that more PS signals are present in the PR interaction. To make the scoring procedure more transparent, we inspected PRs with low, medium, and high PR-level PS index scores (see Table~\ref{tab:pr_examples}). 
These examples illustrate how different combinations of observable PR interaction indicators contribute to the index scores. }

\begin{table}[t]
\centering
\captionsetup{font=footnotesize}
\footnotesize
\caption{\changed{Illustrative PR examples with low, medium, and high PR-level PS index scores.}}
\label{tab:pr_examples}
\begin{tabular}{p{1.8cm} p{0.9cm} p{4.7cm}}
\toprule
\multicolumn{1}{c}{Example} &
\multicolumn{1}{c}{PS index} &
\multicolumn{1}{c}{Observed indicators} \\
\midrule
Low-index PR
& \multicolumn{1}{c}{0/9}
& Few visible interaction signals in the PR thread. \\

Medium-index PR
& \multicolumn{1}{c}{5/9}
& Some discussion, contributor involvement, and direct mention. \\

High-index PR
& \multicolumn{1}{c}{8/9}
& Multiple interaction signals, including visible discussion, participant involvement, and direct mention. \\

\bottomrule
\end{tabular}
\end{table}

We then averaged the PR-level scores for each contributor within a repository, followed by averaging these contributor-level scores to obtain the repository-level index. This gives each contributor equal weight regardless of the number of PRs submitted. The repository-level scores are available in our supplementary material~\cite{figshare}.

\begin{table}[t]
\centering
\captionsetup{font=footnotesize}
\caption{Descriptive statistics for the PS index.}
\label{tab:ps_descriptives}
\scriptsize
\setlength{\tabcolsep}{6pt}
\begin{tabular}{lrrrrrrrr}
\toprule
\multicolumn{1}{c}{Level} &
\multicolumn{1}{c}{N} &
\multicolumn{1}{c}{Mean} &
\multicolumn{1}{c}{SD} &
\multicolumn{1}{c}{Med.} &
\multicolumn{1}{c}{Q1} &
\multicolumn{1}{c}{Q3} &
\multicolumn{1}{c}{Min} &
\multicolumn{1}{c}{Max} \\
\midrule
PR & 60685 & 4.02 & 3.25 & 4.00 &
0.00 & 7.00 & 0.00 & 9.00 \\
Contributor & 14502 & 4.12 & 2.80 & 4.00 &
2.00 & 7.00 & 0.00 & 9.00 \\
Repository & 26 & 3.27 & 1.20 & 3.42 &
2.77 & 4.00 & 0.50 & 5.77 \\
\bottomrule
\end{tabular}
\end{table}

\changed{Table~\ref{tab:ps_descriptives} shows that the PS index captures variation across different levels of analysis. At the PR level, the index uses the full range from 0 to 9, with a mean of 4.02 and a median of 4. This indicates that the score is not concentrated at one end of the scale. At the contributor level, variation is preserved after averaging PR-level scores for each contributor within a repository. At the repository level, the range becomes narrower, but the scores still vary across repositories. 
This shows that the repository-level index captures variation in PR interaction patterns across repositories.}

\section{Survey Study}\label{sec:survey}

\changedd{Alongside the repository-based analysis, we conducted a survey to examine how OSS contributors perceive PS and experience PR interactions. 
While the repository-based analysis constructs a PS index based on PR interaction patterns derived from the proposed framework, the survey provides contributors' perspectives on these aspects of PR collaboration and examines how their reported experiences relate to their perceived PS within the repository.}
 We asked contributors about their perceived PS, their experiences of PR interactions corresponding to the repository indicators, and their broader PR experiences through open-ended questions.\changedd{ The survey received ethical approval from the authors' institution (reference ID: FSE.EC26024)}. This section describes the participant recruitment, survey measures, data analysis, and respondents.

\subsection{Participant Recruitment}
We recruited contributors from the same set of open-source repositories used in the repository-based study (see Section \ref{reposelection}). For contributors who had participated in multiple repositories, we selected the repository in which they had the highest number of PR. This provided each participant with a single project context and prioritized the repository with which they had the most PR experience.  We identified approximately 5,300 potential participants from contributors who had participated in PR activities and for whom a publicly available email address could be obtained.
We contacted the participants by email and invited to complete an online survey about their experiences of PS and PR interactions in the selected repository. The survey took less than 10 minutes to complete. We received 59 completed responses.

As part of
response quality screening, we examined completion status,
response duration, duplicate responses, response patterns (e.g., consistently selecting the same answer option across survey items),
and open-ended responses. All responses were complete,
and no duplicate responses were identified.
We excluded two responses prior to analysis. The first
was identified as a protest response because its open-ended
answers explicitly expressed dissatisfaction with receiving
the survey rather than addressing the study topic. The
second explicitly indicated that it was completed from the
perspective of an AI agent rather than a human contributor.
After these exclusions, the final analytic sample comprised
57 respondents.

\subsection{Survey Measures}
The survey contained two main sets of measures relevant to this study. First, we measured contributors' perceived PS by asking respondents to rate the statement ``Overall, repository \texttt{[repository\_name]} feels psychologically safe for contributors.'' Second, we asked respondents about nine PR interactions corresponding to the signals used in our repository-based analysis (see Table \ref{tab:survey_mapping_results}). Responses were recorded on a five-point Likert scale ranging from ``Strongly disagree'' to ``Strongly agree.''

\subsection{Data Analysis}
We first summarized the response distributions for the nine PR interaction items and the overall PS item. To examine
whether the interaction conditions represented in our repository-based analysis
were associated with contributors' perceived PS, we calculated Spearman's rank correlations between each PR interaction item and respondents' overall reported PS. We used Spearman's correlation because the survey responses were measured on five-point ordinal Likert scales. As nine correlations were tested, we adjusted the resulting $p$-values using the Benjamini-Hochberg~\cite{Benjamini_Hochberg_1995} procedure to control the false discovery rate.

Because several respondents came from the same repository, we additionally calculated 95\% confidence intervals using a repository-cluster bootstrap~\cite{10.1111/j.1467-9868.2007.00593.x} with 2,000 resamples~\cite{Carpenter2000}. This allowed us to check whether the positive relationships remained when accounting for respondents being grouped within the same repository rather than treating all responses as fully independent, as it did with Spearman's rank correlations.

\subsection{Survey Responses}

\subsubsection{Participant Characteristics}
Most respondents were experienced OSS contributors, with
84.2\% reporting more than five years of OSS experience.
The majority identified as contributors (96.5\%), and most
reported submitting (96.5\%) and commenting on PRs (87.7\%). Regarding PR activity frequency, the largest group reported participating sometimes (40.4\%), followed by very often (26.3\%). Most respondents contributed to OSS primarily on a voluntary or personal basis (59.6\%). Based on repository activity, the largest group consisted of one-time contributors (59.6\%). Full participant characteristics are reported in the supplementary material~\cite{figshare}.

\begin{table}[t]
\centering
\captionsetup{font=footnotesize}
\caption{Descriptive statistics for perceived PS
and PR interaction items ($N=57$).}
\label{tab:survey_descriptive}
\scriptsize
\setlength{\tabcolsep}{3pt}
\resizebox{\columnwidth}{!}{%
\begin{tabular}{lrrrrrl}
\hline
Measure & N & Mean &
SD & Median & IQR &
Range \\
\hline
Overall perceived PS
& 57 & 4.28 & 1.06 & 5.00 & 1.00 & 1.00--5.00 \\

Nine-item PR interaction score
& 57 & 4.15 & 0.65 & 4.22 & 0.89 & 2.56--5.00 \\
\hline
\end{tabular}%
}
\end{table}

\subsubsection{Descriptive Statistics}
Overall, respondents reported relatively high levels of perceived PS and positive PR interactions. The overall perceived PS measure had a mean of 4.28 (SD = 1.06, Md = 5.00), with responses ranging from 1 to 5. The nine-item PR interaction score had a mean of 4.15 (SD = 0.65, Md = 4.22). The descriptive statistics are presented in Table~\ref{tab:survey_descriptive}.
\begin{table}[t]
\centering
\captionsetup{font=footnotesize}
\caption{Associations between survey items corresponding to repository variables and perceived PS ($N=57$).}
\label{tab:survey_mapping_results}

\begingroup
\setlength{\tabcolsep}{1.5pt}
\renewcommand{\arraystretch}{1.05}
\scriptsize

\begin{tabularx}{\columnwidth}{
@{}
>{\raggedright\arraybackslash}p{2cm}
>{\raggedright\arraybackslash}X
c c c
@{}
}

\hline
Repository variable &
\multicolumn{1}{c}{Survey item} &
$\boldsymbol{\rho}$ &
BH $p$ &
95\% CI \\
\hline

\texttt{pr\_comment}\newline\texttt{\_num} &
There is enough discussion for contributors to understand feedback and decisions. &
.635 & $<.001$ & [.379,.804] \\

\texttt{has\_exchange} &
Discussions involve back-and-forth communication between contributors and reviewers or maintainers. &
.360 & .009 & [.199,.568] \\

\texttt{contrib}\newline\texttt{\_comment} &
Contributors feel able to respond in pull request discussions. &
.445 & .002 & [.242,.688] \\

\texttt{num\_comments}\newline\texttt{\_con} &
Contributors feel able to ask follow-up questions or explain their choices in detail. &
.330 & .016 & [.207,.571] \\

\texttt{inte\_comment} &
Maintainers or integrators participate in pull request discussions when their input is needed. &
.358 & .009 & [.161,.547] \\

\texttt{reviewer}\newline\texttt{\_comment} &
Reviewers provide comments or feedback during pull request discussions. &
.426 & .003 & [.252,.688] \\

\texttt{other\_comment} &
Other project members join pull request discussions to provide input or support. &
.315 & .019 & [.173,.534] \\

\texttt{num}\newline\texttt{\_participant} &
Discussions involve multiple project members, not only one contributor and one reviewer. &
.369 & .009 & [.197,.697] \\

\texttt{at\_tag} &
Contributors mention other project members to ask for input or clarification. &
.293 & .027 & [.113,.647] \\

\hline
\end{tabularx}

\vspace{1pt}
\parbox{\columnwidth}{\scriptsize
\textit{Note:} $\rho$ denotes Spearman's correlation between each survey item and perceived PS. BH $p$ denotes Benjamini--Hochberg-adjusted $p$-values.
}
\endgroup
\end{table}

\section{Empirical Findings}
\subsection{Associations Between PR Interactions
and Perceived PS}
We examined whether contributors' perceptions of the nine
PR interactions were associated with their overall perceived PS. Table~\ref{tab:survey_mapping_results} presents the mapping between corresponding repository variables, survey items, and their associations with perceived PS.

All nine PR interaction items were positively associated with respondents' overall perceived PS (Table~\ref{tab:survey_mapping_results}). The strongest association
was observed for having enough discussion to understand
feedback and decisions ($\rho = .635$, adjusted $p < .001$),
followed by contributors feeling able to respond
($\rho = .445$, adjusted $p = .002$) and reviewers providing
comments or feedback ($\rho = .426$, adjusted $p = .003$).
All associations remained statistically significant after correction. \changedd{The 95\% confidence intervals were also entirely above zero for all nine items, even after accounting for the possibility that contributors from the same repository share similar collaboration experiences.} 

\changedd{These results suggest that contributors who reported more positive experiences of these PR interactions also tended to report higher perceived PS, and that these relationships remained supported when considering differences in contributors' repository environments. 
The consistently positive associations suggest that the surveyed aspects of PR interactions are relevant to how contributors perceive PS. 
However, these associations are about contributors' reported experiences rather than the corresponding indicators extracted from GitHub. These therefore do not establish that the repository-based indicators directly measure perceived PS.
}

\subsection{Incorporating Contributors' Perspectives into the Construction of PS Index}

The repository-based PS index assigns equal weight to all nine repository indicators (see Section \ref{sec:constructing}). However, the survey results (Table~\ref{tab:survey_mapping_results}) showed that the corresponding PR interactions differed in the strength of their associations with perceived PS. For example, the survey item representing repository variable \texttt{pr\_comment\_num} had a stronger association with perceived PS ($\rho = .635$) than the item representing \texttt{at\_tag} ($\rho = .293$). \changedd{Given that weighting choices can affect the contribution of individual indicators to composite scores~\cite{Becker2017Weights},} this motivated us to examine how giving more weight to indicators whose corresponding survey items had stronger associations would change the repository scores. 

\changedd{Following the broader methodological approach of correlation-based weighting in composite indicators \cite{greco2018}, we conducted an exploratory sensitivity analysis using survey-informed weights.} Specifically, each repository indicator was assigned a weight proportional to the Spearman correlation coefficient between its corresponding survey item and overall perceived PS. 
The weighted score for PR $j$ was calculated as:

\begin{equation}
PS_j^{\mathrm{weighted}} =
\sum_{k=1}^{9}
\frac{9\rho_k}{\sum_{\ell=1}^{9}\rho_\ell} I_{j,k},
\end{equation}

where $\rho_k$ denotes the Spearman correlation for survey item $k$ and $I_{j,k}$ denotes the corresponding binary repository indicator. 
The sum of the nine correlation coefficients is $\sum_{\ell=1}^{9}\rho_\ell=3.531$.
 For example, \texttt{pr\_comment\_num} ($\rho=.635$) receives a weight of $9(0.635)/3.531=1.62$, while \texttt{at\_tag} ($\rho=.293$) receives a weight of $9(0.293)/3.531=0.75$. The weights sum to nine, preserving the index range of 0-9. We then aggregated the weighted PR scores at the contributor and repository levels using the same procedure as the original index.

We found that the weighted scores remained close to the original scores. The largest absolute change was 0.30 points on the 0-9 scale, while the mean absolute difference was 0.08 points. The complete comparison is provided in our supplementary material~\cite{figshare}.
Most scores decreased slightly after weighting but remained close to the original values. 
These findings suggest that incorporating the survey associations into the weights produces similar repository-level scores to the original equal-weighted index. 
\changedd{This suggests that the repository-level scores were relatively insensitive to this alternative weighting scheme, although the similarity does not establish that either index accurately represents contributors' perceived PS.}

\subsection{Correlation Between the PS Index and Contributors' Perceived PS}

We examined whether repositories with higher PS indices also had higher survey-reported PS. To do so, we averaged respondents' perceived PS scores within each repository and compared these averages with the corresponding repository-based indices. Of the 26 repositories in our dataset, 14 had at least one usable survey response and were included in this comparison.
\changedd{We found no statistically significant association between the original repository-based PS index and average survey-reported PS ($\rho = -.182$, $p = .533$). 
However, this finding should be interpreted cautiously. The comparison was based on only 14 repositories, some of which were represented by a single respondent. Furthermore, average survey-reported PS scores ranged from 3.33 to 5.00, with five repositories reaching the maximum score of 5.00. The small number of repositories and limited variation in perceived PS make it difficult to establish whether the two measures are associated. Therefore, although we did not find a statistically significant association, we cannot conclude that the repository-based index is unrelated to contributors' perceived PS.}

\subsection{Exploring PS Index and Contributors' Perceived PS}
To explore how the repository-based PS index and contributors' perceived PS differ, we compared them by classifying repositories into relatively higher or lower PS groups using a separate median split for each measure. This allowed us to compare repositories' relative positions despite the measures having different scales. Repositories at or above the respective median were classified as higher PS, while those below it were classified as lower PS. The medians were approximately 3.62 for the original index, and 4.50 for perceived PS. 
The repository-level scores and classifications for the 14 matched repositories are provided in our supplementary material~\cite{figshare}. 
\changedd{Furthermore, survey-reported PS scores were concentrated toward the higher end of the scale, ranging from 3.33 to 5.00 across repositories. Therefore, these groups should not be interpreted as representing psychologically safe and unsafe repositories.}

\changedd{Given the limited variation in survey-reported PS, we explored whether the index showed a similar higher-lower pattern when comparing repositories at opposite ends of the available perceived PS scores.}
We descriptively compared the mean repository-based PS index of three repositories at the higher end of perceived PS with that of three repositories at the lower end. We selected the first three and last three repositories, ordered by mean perceived PS. 
For the three selected repositories at the higher end of perceived PS, the mean survey score was 5.00 in each repository. At the lower end, the mean survey scores were 4.00, 3.50, and 3.33, giving a group mean of 3.61.
The corresponding repository-based PS indices for the higher-perceived-PS group were 3.81, 4.71, and 3.52, giving a group mean of approximately 4.01. For the lower-perceived-PS group, the original indices were 3.69, 3.20, and 4.06, giving a group mean of 3.65.

For these selected groups, both measures showed the same direction of difference: $5.00 > 3.61$ for perceived PS and $4.01 > 3.65$ for the repository-based PS index. Thus, the group with higher perceived PS also had a higher average repository-based PS index.
We repeated the comparison using all five repositories tied at the highest perceived PS score and the two repositories with the lowest perceived PS scores. Both measures therefore showed the same direction of difference: $5.00 > 3.42$ for perceived PS and $3.71 > 3.63$ for the original index. This again indicates some agreement between the selected group averages.

In other words, the PR patterns summarized by the index reflected the same higher-lower pattern as contributors' reported feelings of PS. This suggests some alignment between the repository-based PS index and contributors' perceptions for these selected groups. However, these findings should be interpreted cautiously. 
\changedd{We compared only a few repositories, some survey PS scores were based on very few respondents, and perceived PS showed limited variation across repositories.}
The two measures showed the same pattern for these selected groups on average, but this gives us limited evidence that they would show similar patterns across repositories more generally.

\subsection{Contributors' Experiences of PS}
We examined the open-ended responses to better understand contributors' experiences and how these relate to our proposed framework. We coded 21 responses describing experiences of PS and eight responses identifying additional conditions that help contributors feel safe. 
\changedd{We used these responses to contextualize the quantitative findings and identify nuances in how contributors experience and interpret PR interactions.
} We present selected responses around the following three themes, with the quotations attributed using anonymized response identifiers (R1, R2, etc.).

\changedd{Respondents described both positive and negative experiences of PR feedback. One respondent who reported high perceived PS (5/5) explained that their input ``was listened to, even if not always agreed with'' (R23), and described the feedback as valuable and clearly explained. In contrast, another respondent who reported lower perceived PS (2/5) described receiving a timely response to their PR but perceived the accompanying closure comment as dismissive (R21). These experiences illustrate how \textbf{contributors' perceptions of PS may depend not only on whether feedback is provided, but also on how it is communicated.}}

Respondents also identified conditions that could support their participation. One emphasized ``patience with differing time availability'' (R11) and the importance of allowing contributors to respond before closing their PRs. Another highlighted the importance of moderation and reporting processes for resolving disagreements (R7). Respondents also mentioned guidance from experienced project members (R3, R20) and recognition of contributors' efforts (R11). These responses suggest that \textbf{opportunities to participate and support from project members may contribute to how psychologically safe contributors feel} when engaging in PR discussions.

Some respondents described delayed responses and unanswered PRs, but interpreted these experiences differently. One respondent considered a six-month delay in acknowledging their PR understandable given the voluntary nature of OSS (R15). Another described having their proposal rejected but considered the reviewer fair and polite (R18). Similarly, a respondent whose contributions were overlooked or modified without explanation stated: ``I didn't feel unsafe, but I also didn't feel like my contributions were particularly valued'' (R24). These responses illustrate that PR outcomes do not necessarily reflect contributors' perceived PS, as their \textbf{interpretation also depends on how interactions are handled and the circumstances surrounding them}.

\section{Discussion}\label{sec:VI}

\subsection{Interpreting the Framework Through Two Perspectives}

\changedd{Our mixed-methods findings provide different perspectives on how PS may be reflected in PR interactions. The repository-based analysis demonstrates how selected aspects of the proposed framework can be represented through observable patterns of PR discussion, participation, and requests for input. However, the occurrence of these interactions does not necessarily reveal how contributors experience them. 
For example, a high volume of PR comments may reflect constructive discussion, but previous studies have also linked increased discussion to technical complexity and confusion during code review \cite{azeem2024codesmells,ebert2021confusion}. Similarly, direct mentions may reflect the need to involve relevant contributors \cite{kavaler2019mentions}, while maintainer participation may be influenced by project governance practices \cite{alami2021pull}.
This is consistent with Ferguson et al.'s study of PS characteristics in online communication, which identified several potentially useful observable signals while acknowledging the limitations of automated measurement \cite{ferguson2024no}. }

\changedd{The survey provides another perspective by examining contributors' reported experiences of PR interactions and their perceived PS. All nine surveyed interaction items were positively associated with perceived PS, which suggest that the aspects identified through our framework are relevant to how contributors perceive PS. These findings are consistent with previous research about the importance of speaking up, communication, and collaborative behavior in psychologically safe software teams \cite{Santana2023,Alami2024,sanchezgordon2026voice}. Contributors' open-ended responses further showed nuances of how PS can be reflected in PR interactions, such as feedback quality, feeling heard, and the interpretation of PR outcomes. Similarly, Turzo and Bosu found that the usefulness of code review comments depends not only on technical contributions but also on their comprehensibility and politeness \cite{Turzo2023}.} 


\changedd{Both approaches highlight the difference between observable PR interaction patterns and contributors' perceptions of PS. Although we observed no statistically significant association between the repository-based index and survey-reported PS, the comparison involved only 14 repositories with limited variation in perceived PS. Therefore, this result does not establish whether the two are generally associated. However, incorporating survey-informed weights produced only small changes in the repository-based index scores. Furthermore, our exploratory comparisons showed that repositories with higher average perceived PS also had higher average index scores. These findings suggest some similarities between the two measures for selected groups of repositories, but do not establish that the index can reliably distinguish repositories based on contributors' perceived PS.}


\changedd{Nevertheless, the qualitative findings illustrate why the presence of observable interactions alone may not fully represent contributors' experiences. For example, receiving feedback may provide an opportunity for discussion, but whether contributors feel psychologically safe may also depend on how that feedback is communicated. This distinction is important when considering whether observable indicators adequately represent a theoretical construct \cite{SjobergBergersen2023}.}


\changedd{Taken together, our findings suggest that the proposed framework offers a basis for examining PS-related aspects of PR collaboration through different sources of evidence. Repository-based indicators make selected framework items observable, while contributors' reported experiences provide insight into their perceived PS and the nuances surrounding these interactions. 
While each approach captures different aspects of PS-related collaboration, together they provide an initial, more nuanced understanding of how PS may be reflected in PR interactions within pull-based OSS projects.
}


\subsection{Implications For OSS maintainers and contributors}
Our findings highlight the importance of considering not only how often contributors interact during PR discussions, but also how these interactions are conducted. Contributors who reported having enough discussion to understand feedback and decisions, opportunities to respond, and reviewer involvement also tended to report higher perceived PS. Previous studies have highlighted the importance of useful code review feedback \cite{Bosu2015}, including the comprehensibility and politeness of review comments \cite{Turzo2023}. Furthermore, Alami et al.~\cite{Alami2024} found that PS in agile software teams is associated with behaviors such as speaking up, helping others, and collective problem-solving. Based on our findings and these studies, maintainers could pay more attention in providing clear and constructive feedback, to allow contributors to explain their decisions, and handling disagreements respectfully. However, delayed responses or rejected PRs do not necessarily make contributors feel unsafe, as their experiences also depend on the circumstances surrounding these interactions.

\section{Limitations}

Several limitations should be considered when interpreting our findings. 

First, the two empirical approaches capture different aspects of PR collaboration and perceived PS. The repository-based analysis relies on observable PR interaction data and examines only selected parts of the proposed framework. Its indicators do not directly capture contributors' feelings or the quality and context of their interactions. The survey, on the other hand, captures contributors' reported perceptions and experiences, but these cannot be directly linked to their observable PR interactions. Therefore, the two approaches should not be treated as interchangeable measures of PS.


Second, the construction of the PS index necessarily simplifies complex PR interactions. The PR-level index is based on binary indicators and gives each indicator equal weight. While this makes the index transparent and reproducible, it may not fully reflect the relative importance of different indicators. To examine the influence of equal weighting, we conducted a sensitivity analysis using survey-informed weights. The resulting scores remained relatively similar.

Third, our survey findings are subject to sampling and measurement limitations. We recruited contributors whose email addresses were publicly available, and participation was voluntary. Consequently, our 57 respondents may not represent the broader population of OSS contributors. To improve response quality, we screened responses for completion, duration, duplicates, and response patterns, and excluded two responses that did not meet our inclusion criteria.

Fourth, comparing the repository-based index with survey-reported PS has several limitations. The repository data and survey responses reflect different periods, during which project practices and contributors' experiences may have changed. Furthermore, we could only compare the two approaches at the repository level rather than link individual contributors' perceptions to their actual PR interactions. The comparison was also limited to 14 repositories, some represented by only one respondent. Restricting the analysis to the 11 repositories with at least two respondents did not change the significance of the result.

Finally, the framework and empirical study are grounded in GitHub PR data from 26 highly starred repositories. OSS collaboration may also take place through other channels, such as issues, mailing lists, chat platforms, or private communication, which are not captured in our analysis. In addition, highly starred repositories may have different review norms, governance structures, and contributor dynamics compared with smaller or less visible OSS projects. Future work could apply and refine the framework across a broader range of OSS contexts and communication channels.


\section{Conclusion}


In this study, we proposed a theory-informed framework for understanding how psychological safety (PS) may be reflected in pull request (PR) interactions in open-source software projects. \changedd{We investigated the framework through a mixed-methods approach combining repository-based analysis of 60,685 PRs across 26 GitHub repositories and a survey of 57 OSS contributors.
The repository-based analysis constructed the PS index using the selected observable patterns of PR interactions, while the survey showed that all nine interaction items surveyed were positively associated with contributors' perceived PS. However, we found no statistically significant association between the repository-based index and survey-reported PS. Contributors' open-ended responses further highlighted the importance of feedback quality, opportunities to participate, and the circumstances surrounding PR interactions.
Together, these findings suggest that observable PR interaction patterns and contributors' reported experiences provide different perspectives on PS. Our framework offers a basis for examining these aspects in pull-based OSS projects, while also highlighting the importance of considering contributors' experiences when interpreting repository PR indicators.}

\section*{Data Availability}
\label{sec:data}
Our supplementary materials \cite{figshare} contain source code used to retrieve and analyze data, results for all our analyses, and instructions how to replicate the results.

\bibliographystyle{IEEEtran}
\bibliography{conference_101719}

\end{document}